# Pulsatile therapy for perovskite solar cells

*Kiwan Jeong[1†], Junseop Byeon[1,2†], Jihun Jang[1], Namyoung Ahn[1]\*, and Mansoo Choi[1,2]\**

[1]Global Frontier Center for Multiscale Energy Systems, Seoul National University, Seoul 08826, Republic of Korea.

[2]Department of Mechanical Engineering, Seoul National University, Seoul 08826, Republic of Korea.

\* To whom correspondence should be addressed.

E-mail: Mansoo Choi (mchoi@snu.ac.kr), Namyoung Ahn (nyny92@snu.ac.kr)

† These authors contributed equally to this work.

## Abstract

The current utmost challenge for commercialization of perovskite solar cells is to ensure long-term operation stability. Here, we developed the pulsatile therapy which can prolong device lifetime by addressing accumulation of both charges and ions in the middle of maximum power point tracking (MPPT). In the technique, reverse biases are repeatedly applied for a very short time without any pause of operation, leading to stabilization of the working device. The observed efficacies of our pulsatile therapy are delaying irreversible degradation as well as restoring degraded photocurrent during MPPT operation. We suggest an integrated mechanism underlying the therapy, in which harmful deep-level defects can be prevented to form and already formed defects can be cured by driving charge-state transition. We demonstrated the therapy to maintain defect-tolerance continuously, leading to outstanding improvement of lifetime and harvesting power. The unique technique will open up new possibility to commercialize perovskite materials into a real market.

As photovoltaics employing hybrid perovskite halides have continuously been breaking world-records of power conversion efficiency (PCE), expectations for their industrialization are rapidly rising(*1-7*). Perovskites can be utilized for a wide range of applications like flexible and wearable power sources(*8, 9*), tandem devices with Si solar cells(*10*), also other photonic devices such as light emitting diodes(*10, 11*) and photodetectors(*12, 13*). However, real markets will require insurance for a long-term operational stability of perovskite devices and unfortunately, the insurance of lifetime has not been guaranteed yet(*14*) . The current level on device stability is just about few thousands of hours under one sun light illumination and the understanding on fundamental mechanism of perovskite device degradation is still limited.

As part of an effort to solve this stability issue of perovskite solar cells, many studies were performed. Degradation studies picked up ion instabilities and trapped charges as the main reason of perovskite instabilities(*11, 15-17*). Ions are easily migrated through defect sites by electric field and light illumination, leading to material transformation(*18*). Although several reports showed a positive role of ion migration like defect healing (*19*), most of relevant studies demonstrated instabilities induced by halide migration and segregation(*20-23*). Interestingly, it was confirmed that degradation induced by ion instabilities occurred somewhat reversibly. Decreased performance was recovered back after storing the degraded device in the dark chamber. But, the recovery took several hours (usually overnight) for spontaneous ion redistribution.

Carrier charges trapped in perovskite films were found to induce irreversible degradation in the presence of $H_2O$ and/or $O_2$, producing stable yellow lead(II) iodide and lead hydroxide species (*17*). Detailed chemical routes of trapped-charge driven degradation were identified via *ab initio* molecular dynamics simulation(*24, 25*). Oxygen driven degradation studies also suggested an important role of localized charges in the formation process of superoxides(*25, 26*). It was recently confirmed that such charges play a critical role in irreversible device-level degradation(*17, 27*). Ion migration instability was also found to be deeply involved with localized charge trapping(*28*). We can learn from all of the studies that both trapped charges (charge trapping) and halide defects (ion migration process) are mainly responsible for perovskite device instability. Thus, to achieve the long-term stable perovskite solar cells, we have to find a new strategy to effectively control two fundamental reasons.

Herein, we developed a novel way (called pulsatile therapy) to lengthen device lifetime by effectively extracting accumulated charges in the middle of maximum power point tracking(MPPT). In the therapy, a pulsatile reverse pulse(RP) bias is repeatedly applied for a very short time to eliminate charges as well as drift ions reversely, leading to stabilization of the working device. To systemize our pulsatile therapy (PT), we built a new pulsed-MPPT system that can apply desired rectangular pulse waves of bias voltage, in which amplitudes for both MPPT and RP were programmably calculated from feedback information of actual *J-V* characteristics of the target cell. The system demonstrated that the PT significantly improved total power harvesting and device lifetime in comparison to the normal MPPT case for different types of perovskite solar cells, though not fully optimized yet. Strikingly, it was clearly observed that the therapy effects were not only to make device degradation slower, but to restore degraded power shortly after a RP.

To further elucidate the mechanism underlying our PT, we studied photoluminescence (PL) changes during the therapy together with photocurrent(PC) dynamics. We confirmed charge extraction and ion redistribution from two exponential components of PC decay in the reverse bias step, which also accompanied steady rise of PL intensities indicating defect healing. In the following MPPT step, both PL and PC signals restored from their degraded values again, which is indicative of reduction of non-radiative recombination. Based on all observations, we suggested an integrated theory on degradation and recovery which is deeply involved with charges, ions, and defects. Concentration of shallow traps ($I_i^{-1}$ and $Pb_i^{+2}$) and trapped charges increase during degradation (at MPPT), thereby charge-state transition actively happens to form deep traps of interstitial defects ($I_i^0$ and $Pb_i^0$). The following RP of our therapy returns the deep traps to shallow traps by extracting charges and may further induce defect annihilation through drifting ions reversely. Such a defect-healing process restores and stabilizes the degraded device again, delaying irreversible degradation and pulling performance up reversibly. Based on the nature of defect-related degradation, we discovered better PT to ultimately prevent the formation of deep traps by reducing time for charge-state transition, leading to improvement of total power generation and lifetime. This technique is the first method to heal the device without pause of operation and prolong device lifetime.

**Dependency of electrical operation conditions on device degradation**

All types of photovoltaic devices always experience performance degradation with aging. But, their lifetime will vary greatly depending on photostability of a photoactive material. Unfortunately, photostability of perovskite materials are still very deficient to withstand continuous light illumination. To tackle the stability issue, our group has studied and investigated the origin of instability, from which it has been confirmed trapping of photogenerated charges is mainly responsible for light-induced instability of perovskite materials(*17,24,25,29*). Together with charges, ion instabilities of these materials play a critical role in degradation (*21,30*). However, there still exists no model integrated with two main reasons (charge trapping and ion instabilities) to account for their interplay to cause fast degradation. Later this paper, based on new experimental results observed from our pulsatile therapy, we will suggest an integrated theory in which both charges and ions are involved.

To begin with, we investigated photo-stability of three different devices by measuring time evolution of normalized power under one sun illumination at three different electrical conditions (short-circuit (SC), maximum power point tracking (MPPT), and open-circuit (OC), respectively). (See **Fig. 1A**) Detailed experimental results and explanations about those three devices are in **fig. S1** and Materials and Methods of Supplementary Materials. In all three devices, the fastest degradation was observed when the device was kept open-circuited. The other two states, MPPT and SC, showed much slower degradation rates as compared to the OC cases. SC conditions led to a slightly better stability than MPPT conditions. Such bias dependency of the degradation speed was also previously confirmed in previous degradation studies (*31,32*). We additionally performed photo-stability test at four different bias around initial maximum power(MP) point as shown in **Fig. 1B**. The results again revealed that higher bias voltage caused faster degradation. *J-V* characteristics of devices biased at lower voltage than 0.65V was not even changed for 100 hours. (**fig. S2**) Those are clear evidence of interplay between charges and ions because the bias voltage accumulates photo-generated charges as well as activates migration of charged ions (vacancies) by electrostatic coulombic force. (see **Fig. 1C**) Moreover, it was previously confirmed that localized charges induce ion migration and segregation*(28)*, which is evidence to show their interplay causing material instabilities. Namely, at higher bias voltages, perovskites must be degraded rapidly due to the combination of two major causes (*31,32*).

**A new system demonstrating pulsatile therapy and its efficacies**

An interesting observation regarding device degradation is that degradation rates at MPPT also depend on time interval of *J-V* sweeps as presented in **fig. S3**. The shorter time interval (5 min) led to better stability than the longer interval (30 min). Those imply not only that the duration of MPPT operation may affect the operation stability, but that *J-V* sweeps may also slow down device degradation by releasing accumulated charges and ions. Electrical operation conditions, especially for these sensitive perovskites, play a critical role in device degradation as confirmed above. In the present study, focusing on an electrical technique to ease two fundamental reasons, we developed a pulsed-MPPT system to harvest more power and lengthen device lifespan which is practically applicable without stopping operation. To extract accumulated charges during a period of MPPT working, a short reverse pulse (RP) bias is applied periodically by our programmed algorithm in this new system. (See **Fig. 2A**) Our system was specially developed to provide a fully automatic pulsed-MPPT by sending and receiving continuous feedbacks between tested devices and the system. (See details in Experimental Section) A specific value of applied RP bias is determined to eliminate potential difference across a perovskite intrinsic layer based on our simple circuit model including accumulated charges (**Supplementary Note 1**) and updated periodically using photovoltaic parameters calculated from our software, thereby we can obtain optimal pulse amplitude for the novel tracking system.

**Fig. 2B** shows an applied pulse wave of bias voltage and corresponding photocurrent (PC) values over time of a device operated by this system. *J-V* curves are obtained periodically to estimate photovoltaic performance of the tested device. As mentioned before, this system uses recently-updated *J-V* characteristics to calculate MP and pulse bias. During power harvesting, applied bias over time will be a simple pulse wave (rectangular wave) that has an amplitude of maximum power voltage($V_{MP}$) in a forward bias. In the RP process with an amplitude of $V_{RP}$, electrical power will be consumed to apply RP, but energy loss will be far lower than energy harvesting as weak reverse biases are applied for a shorter time than energy harvesting process. Energy losses occurred during RP are strictly included for precise comparison. We call this tracking method as pulsatile therapy (PT) from now on. To fairly evaluate the efficacy of PT, we prepared four cells in one ITO-patterned glass that are electrically independent by additional ITO etching. (see Materials and Methods of Supplementary Materials) As each cell was fabricated through the same process on one substrate, all these cells were very similar in terms of performance and degradation. It was confirmed from the fact that these four cells showed similar photovoltaic performance and degradation rates when they were tested in the same condition as shown in **fig. S4**.

We tested the efficacy of PT for Cs-doped $FA_{0.92}MA_{0.08}PbI_{3-x}Br_x$ perovskite solar cells with high efficiency of 20% (ITO/ $SnO_2$/perovskite/Spiro-MeOTAD/Au). In this PT test, its time width for MPPT($T_{MP}$) and RP($T_{RP}$) was fixed at 30 minutes and 30 seconds, respectively. General MPPT was simultaneously performed for the other identical cell in the device. **Fig. 2C** shows time evolution of normalized power for PT- and MPPT- tested cells. While normalized power of the cell with PT maintained 92.5% of the initial power during 40 hours, that of the MPPT-cell decrease by 90.3% of the initial power. To find out how the efficacy occurred, we observed kinetics of normalized power before and after a RP after 2 hours and 16 hours. (see **Figs. 2D and 2E**) Although normalized power linearly decreased during two pulse periods, the decay speed of the PT was slower than that of the MPPT cell. Namely, the PT slows degradation down, which is the first efficacy of the PT as shown in **Fig. 2D**. After some progression of degradation (16 hours), power was dramatically restored and exceeded normalized power value of previous MPPT shortly after a RP as can be seen **Fig. 2E**, which was not observed in earlier times . The PT-induced recovery looks similar to self-healing in dark conditions (*33-35*). Note that our PT can restore degraded performance in just 30 seconds without pause of operation. This recovery is another efficacy of our PT. Those consequently result in improved power generation and longer lifetime, which is the ultimate goal of this technique.

As clinical trials on the PT, pure MAPbI$_3$ solar cell employing vapor-deposited C$_{60}$ ETL was also tested in the same way. (see **fig. S5**) Interestingly, improvement induced by PT was consistently seen The kinetics of normalized power over time were similarly observed as shown in **fig. S5**, which clearly showed the recovery of power after RPs. That means the PT can work with different types of perovskite solar cells even though the efficacies can be somewhat different. These can be simply understood from the fact that timescales of charge trapping and ion movement are varied depending on types of perovskite and charge-transporting layer (*36*). We also confirmed cross-sectional SEM images of the degraded devices operated by PT and MPPT for 50 hours, respectively.(see **fig. S6**). These results clearly revealed that the PT led to delaying irreversible chemical decomposition by mitigating accumulation of charges and ions compared to the conventional MPPT case.

**Mechanism of pulsatile therapy**

Photoluminescence (PL) provides decisive information about charge carriers and defects. PL intensities, which is corresponding to how much radiative recombination occurs, are a critical indicator of quality of a perovskite film (*37*). Especially, PL signals emitted from a device can reflect its photovoltaic performance. It was previously confirmed that decrease of PL intensities accompanied device gradation, which was interpreted as the formation of nonradiative recombination centers by degradation(*30,38*). To obtain PL information of our PT device, we customized PL setup to detect time evolution of PL emission from the device controlled by pulse waves of voltage bias. (see **Fig. 3A**) The desired voltage bias can be applied by a LabVIEW-controlled electro source meter (Keithley 2400) and PC flowing from the measured device will be recorded at the same time. Continuous irradiation of 150W Xe lamp was used to excite our sample. (The intensity of the light source is around 75 mW/cm$^2$)

We observed steady-state PL of our Cs-doped FA$_{0.9}$MA$_{0.1}$PbI$_{3-x}$Br$_x$ device under different bias voltage (**fig. S7**). The peak intensities got higher as we increased the bias voltage because of augmented carrier densities. PL intensity changes at the peak of 778 nm under both forward and reverse voltage sweep were recorded as shown in **fig. S7B**. The *PL-V* curves, which is similar to the *J-V* curves, clearly show the effect of bias voltage on carrier densities (radiative recombination) in the device. Those results also mean our PL measurements work well for the device under electrical bias. We firstly investigated kinetics of PL and PC while the device was operating on MPPT condition for over 1 hr as shown in **Fig. 3B**. Please note that the extremely steep decay of PL and PC signals appearing early in the measurements originated from transient change of bias voltage and light illumination. After that, PL intensities decreased, at the same time, PC values decreased. PL intensity decrease can be explained by augmentation of nonradiative recombination centers(*30,38*), which can account for accompanied PC (performance) degradation. That is to say, the photoactive material underwent the formation of deep-level defects causing nonradiative recombination. In the early stage before 1000 seconds, PC decrease took place more badly than PL, which can be attributed to ion accumulation to form a barrier of charge extraction(*39*).

To rigorously examine how the PT works, we detected time-evolving PL intensity and PC for 30 seconds during a RP applied after 30 minutes of MPPT operation. (see **Fig. 3C**) PC dynamics induced by the RP has two exponential decays, of which the fast one (0.593 sec of time constant) results from capacitive current and the slow one (34.23 sec of time constant) indicates field-induced ion movement(*40*). (see **Table 1**) The results imply that the RP leads to charge extraction and ion redistribution. The more intriguing point is that PL intensities gradually increase by the RP, which would be indicative of the annihilation of defects. **Fig. 3D** shows overall kinetics of PL intensity and PC during one cycle of PT (MP→RP→MP). We confirmed that saturated values of both PL intensity and PC restored and exceeded their values that were degraded during the previous MP process (see red arrows in **Fig. 3D**), which evidently explains restoration of normalized power as well as reduction of nonradiative recombination. (see **Fig. 2E**) (Please note again that transient changes in PL intensity and PC originate from cascade change of bias voltage (RP→MP)). Together with the gradual rise of PL intensities during the RP step, the recovery of PL intensity provides obvious evidence of defect annihilation. The kinetics were confirmed reproducibly as can be seen in **fig. S8**. The short RP effectively extracts carrier charges and releases ion accumulation without any pause of operation, leading to the defect annihilation and the resulting reduction of nonradiative recombination. As a consequence of RP-induced transformation, the device can be stabilized again after the pulse, thereby irreversible degradation may be delayed and reversible recovery occurred (see **Figs. 2D and 2E**).

One unique character of the perovskite is defect-tolerance, which means that intrinsic defects are hardly harmful for performance because they are located on intra-band or shallow states(*41,42*). So, they cannot act as deep-level traps or nonradiative recombination centers that are detrimental to device performance(*43,44*). Defect tolerance is mainly why perovskite solar cells show high performance even though we just employed polycrystalline films with lots of grain boundaries(*41,42*). But, from the stability point of view, polycrystalline perovskite films will be disadvantageous due to rich defect densities at grain boundaries. In the process of degradation, such defects can be transformed into harmful deep-level traps through charge-state transition. After all, the degraded perovskite will be no longer defect-tolerant due to the formation of deep-level trap states. This transition must be deeply associated with device performance and stability. In our PT, these harmful defects will reversibly return to harmless shallow defects or ultimately the annihilation of defects by extracting trapped charges and alleviating ion accumulation. As a result, performance is instantaneously recovered due to decrease of nonradiative recombination. More than slight recovery of performance, irreversible degradation slows down a lot owing to the reduction of trapped charges in deep levels.

We propose a possible mechanism underlying the PT based on dynamics of lead and iodine interstitial defects. (see **Fig. 3E**) Note that the PT shows the therapeutic effect in both MA-rich and FA-rich cases. The organic cations and their defects may fairly interplay, but, in this study, we focus on lead cation, halide anions, and the relevant defects because those are more responsible for deep-level trap states than organic cations(*45*). A freshly-fabricated perovskite film already possesses vacancies and some interstitials which have very low formation energies. Most are placed within intra-bands, and the rest occupy shallow-level states close to the band edges. If a perovskite device starts to operate under light illumination at MP, charge carriers, ions, and defects in a perovskite film will start to move by drift-diffusion model(*46*). At the same time, Frenkel and Schottky pairs can be formed by external energy, which means that total defect densities increase during operation, resulting in easy ion migration through vacancies.(*30*). When equilibrium is reached, their concentrations are no longer uniform across the device. At interfaces, high concentrations of carriers and ions will be observed to compensate internal electric field. For example, $I_i^-$ will move toward a hole-transporting layer, whereas $Pb_i^{2+}$ (or $I_i^+$) will go to an electron-transporting layer and accumulate there. Iodine anion will move more aggressively due to its low activation energy of migration. *(18)* Meanwhile, to align fermi energy level with the hole- (or electron-) transporting layer, hole (or electron) carrier concentrations near the interface increase, finally resulting in fermi energy level shift toward valence (or conduction) band edge. In this condition, the formation energies of $I_i^0$ and $Pb_i^0$ become lower than those of $I_i^{-1}$ and $Pb_i^{2+}$*(41)*, which means that charge-state transition can energetically be available as the following reaction.

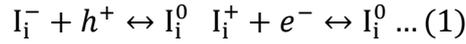
$$I_i^- + h^+ \leftrightarrow I_i^0 \quad I_i^+ + e^- \leftrightarrow I_i^0 \ldots (1)$$

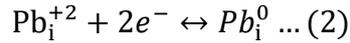
$$Pb_i^{+2} + 2e^- \leftrightarrow Pb_i^0 \ldots (2)$$

Such reactions happen slowly, but steadily during MP operation. If overall carrier densities increase, for example under strong light illumination or forward bias, these reactions will happen more rapidly. *(30)* The neutral interstitial defects occupy deep-level states within bandgap, which is why non-radiative recombination increased (PL intensity and PC decreased) during MP operation. *(30, 47)* (see **Fig. 3b**) These harmful defects will not only trap carriers for a long time, but also cause further chemical reactions to form $I_2$ and $PbI_2$*(30, 48)*. Note that irreversible chemical decomposition occurs in the presence of long-lived charges. *(17)* Stable products like $PbI_2$ cannot return back (irreversible degradation), which is why deep-level defects for long-lived charges are critical to degradation speed. Interestingly, $MA_i^0$ or $FA_i^0$ can be formed in a similar way, but the states of those defects are still shallow. (45) They will not significantly affect PL intensity and PC.

If a RP is instantaneously applied during our PT, carriers and ions will reversely move from the previous MP step. As a result, the charge-state transition will occur adversely because the neutral interstitial defects are no longer energetically favourable when electrons or holes vanish out by the RP. (see above **reaction 1 and 2**) Therefore, deep-level defects gradually vanish out, leading to the rise of PL intensities. (see **Fig. 3C**) In this process, perovskite becomes defect-tolerant again and we might be able to expect the annihilation of defects through recombination of vacancies and interstitials.*(30, 49)* . When the device comes back to MP condition, the healed perovskite film shows better performance with the aid of reduced nonradiative recombination. (see **Figs.2E and 3D**) Long-lived charges will be also reduced again, which makes its irreversible degradation slower based on previously reported trapped charge driven degradation mechanism(*17*).

**Fig. 3E** summarizes the process and efficacy of our PT. As mentioned above, performance degradation occurring during MP operation accompanies material transition from defect tolerance to intolerance as a result of the formation of deep-level defects such as $I_i^0$ and $Pb_i^0$.*(45)* However, the PT effectively induces the healing of deep-level defects using a short RP, ultimately stabilizing the device into the defect-tolerance state with instantaneous recovery of performance. Please note that the PT cannot perfectly prevent degradation in this current device because external reactive chemicals such as water and oxygen can cause irreversible degradation in the presence of accumulated charges (*17*). Besides, there still exist performance degradation of charge-transporting layers and electrodes(*50, 51*). These are possibly why irreversible performance degradation still occurred even if the device was fresh in the defect-tolerance state. It is noteworthy that the recovery of normalized power hardly appeared at the beginning stage and became gradually prominent as time went on. (see **Figs. 2D and 2E** and **fig. S5**) These results indicate that the perovskite film remained defect-tolerant at the beginning stage of degradation (irreversible degradation only occurred) and later became defect-intolerant by possessing deep-level defects as degradation proceeded (both irreversible and reversible degradation occurred).

**Pulsatile therapy holding defect-tolerance for a long time**

One important efficacy of the PT is slowing the degradation speed down by extracting accumulated charges even if the device is still on defect-tolerance. (see **Fig. 2D**) If the PT can deter the transition from defect tolerance to intolerance by eliminating two sources more frequently, we can expect longer lifetime of the device because it has no harmful defects. In this case of defect-tolerance state, there is no reversible recovery. Based on this concept, we optimized our PT for FA-based cells, in which a RP was applied every 10 minutes for 2 seconds. The reason why we chose just 2 seconds for the RP was not only to reduce energy loss during the pulse, but 2 seconds being sufficient to extract charges as shown in **fig. S9**. It was confirmed from charge extraction measurements that most of accumulated charges were extracted within 2 seconds. **Fig. 4A** shows different efficacies of PT depending on time width of pulse. The detailed time dependent power profile including transient signal from PT is shown in **fig. S10A**. The new PT condition ($T_{MP}$=10min $T_{RP}$=2sec) led to more improved stability than the previous PT case ($T_{MP}$=30min $T_{RP}$=30sec). In the new PT, 94.7% of the initial power maintained after 40 hours of continuous operation, whereas the MPPT-cell degraded by 90.3% of its initial power. For FA-rich perovskite solar cells, the new PT showed the efficacies in terms of lifetime and total energy harvesting. We calculated how much the total energy harvested by the PT improved over that harvested by the MPPT, in which energy consumption during reverse pulses was strictly considered. (**Fig. 4B**) Since their initial PCEs were very similar, but slightly different, we used normalized power to estimate normalized total gain of the PT. The detailed calculation steps are included in **Normalized total gain of Supplementary Text**. According to the calculation formula, we can simply consider our PT effective when the normalized total gain starts to exceed 0%. In **Fig. 4B**, both cases showed that the values decreased at the beginning of operation due to RP-induced power loss. Remarkably, the new condition ($T_{MP}$=10min $T_{RP}$=2sec) rapidly exceeded 0% just after 2 hours of operation. On the other hand, the previous condition, which showed reversible recovery shortly after reverse pulse, did not exceed the threshold (0%) during 40 hours owing to significant electrical power losses by RPs even though the lifetime (normalized power) itself was improved (**Fig. 4A**). We confirmed that the previous condition finally crossed the threshold when tested for 45 hours as shown in **fig. S11**. Obviously, the enhanced efficacy of the new PT originated from lower power consumption during reverse pulses. More importantly, there is a remarkable efficacy of the new PT which can prevent the formation of harmful defects for a long time by extracting accumulated charges.

**Fig. 4C** shows kinetics of normalized power over time for the new PT condition, in which no power recovery appeared even after 40 hours of continuous operation. To directly compare two PT conditions in terms of power kinetics, we obtained power recovery occurring after RPs as a function of operation time as shown in **Fig. 4D**. It is interesting that no power recovery was observed for total 40 hours in the new PT test, while under the previous PT condition, power recovery began to be observed after about 10 hours of operation. Kinetics of normalized power over time for the previous PT ($T_{MP}$=30min $T_{RP}$=30sec) was shown in **fig. S10B**. Since the reversible recovery after RPs was observed when harmful defects were formed during continuous MP operation, these results clearly indicate that the new PT case effectively deterred the formation of harmful defects (transition from defect-tolerance to intolerance) by extracting charges and controlling ions every 10 minutes. Although power recovery is definitely a good signal as evidence of defect healing, maintaining defect-tolerance (with no harmful defects) is the best for device stability as harmful defects provide potential irreversible degradation sites. Namely, the ultimate goal of our PT is delaying irreversible chemical decomposition caused by charge trapping, especially like long-lived traps in deep-level states. The new PT condition ($T_{MP}$ =10min $T_{RP}$=2sec) is a better therapy than the previous one since it could prevent the formation of deep-level defects and improve device stability by effectively extracting charges and controlling ions. We also confirmed the reproducibility of new PT as presented in **Fig. 4E**. The PT always shows off the efficacy of lifetime improvement. As each device has inherently different lifetime, test results are not exactly the same. In the one device, different two cells showed very similar results when both were tested under the same PT. (see **fig. S12**) Note that our PT is flexible in choosing $V_{RP}$, $T_{RP}$, and $T_{MP}$ and can be further optimized for best results depending on the type of perovskite solar cells. Our PT will be a reliable and applicable technique to address the stability issue of perovskite solar cells.

In conclusion, our PT is the first technology to cure degraded perovskite solar cells by applying a short electrical pulse of RP without any pause of operation. We confirmed that the therapy can effectively stabilize damaged devices again by extracting charges and redistributing ions. Both MA-pure perovskites and FA-rich perovskites were clinically tested, showing similar efficacies that led to slowing down their degradation speed as well as recovering degraded power reversibly. Our PT shares its special efficacies with those by self-healing in dark, but appears rapidly just in a few seconds, which makes itself energy-friendly. We suggested the mechanism underlying our PT that can account for device degradation and recovery, in which interstitial defects play a decisive role in both reversible and irreversible degradation. Especially, neutral lead and iodine interstitials, which are formed via charge-state transition due to charge trapping and ion accumulation occurring at interfaces during solar cell operation, behave as deep-level trap states, causing performance degradation. Remarkably, the therapy heals those harmful defects by stabilizing both charges and ions, thereby it achieves the defect-tolerance state, which was evidenced by PL intensity rise after a RP. Taking advantage of the defect nature, we optimized the therapy condition to prevent material transformation from defect tolerance to intolerance. In the condition, although there existed no reversible recovery, the efficacies were more effective in terms of lifetime and total power harvesting as compared to the previous one. This study suggests a new mechanism of degradation and recovery based on defect dynamics and extraction of charges and opens up new approach to heal devices quickly and energy friendly. With this technique, possibility of perovskite solar cells to enter a real market will be raised to the full.

## Acknowledgements

We sincerely thank G. Min ,Y. Jeong, and Professor J. Ha for valuable discussion and support to design our PT systems. **Funding**: This work was supported by the Global Frontier R&D Program of the Center for Multiscale Energy Systems funded by the National Research Foundation under the Ministry of Education, Science and Technology, Korea (2012M3A6A7054855). **Author contributions**: M.C. and N.A. conceived the idea of the work. N.A. and M.C. developed a theory by discussing with K.J. and J.B. K.J and J.B. contributed equally to the work. K.J., J.B and N.A. conducted experiments. J.J. did substrate preparation.

N.A., M.C., K.J. and J.B. designed experiments, analyzed data, participated in discussion and wrote the manuscript. M.C. led the work. **Competing interests**: None declared. **Data and materials availability**: The data that support the findings of this study are available from M.C. and N.A. upon reasonable request.

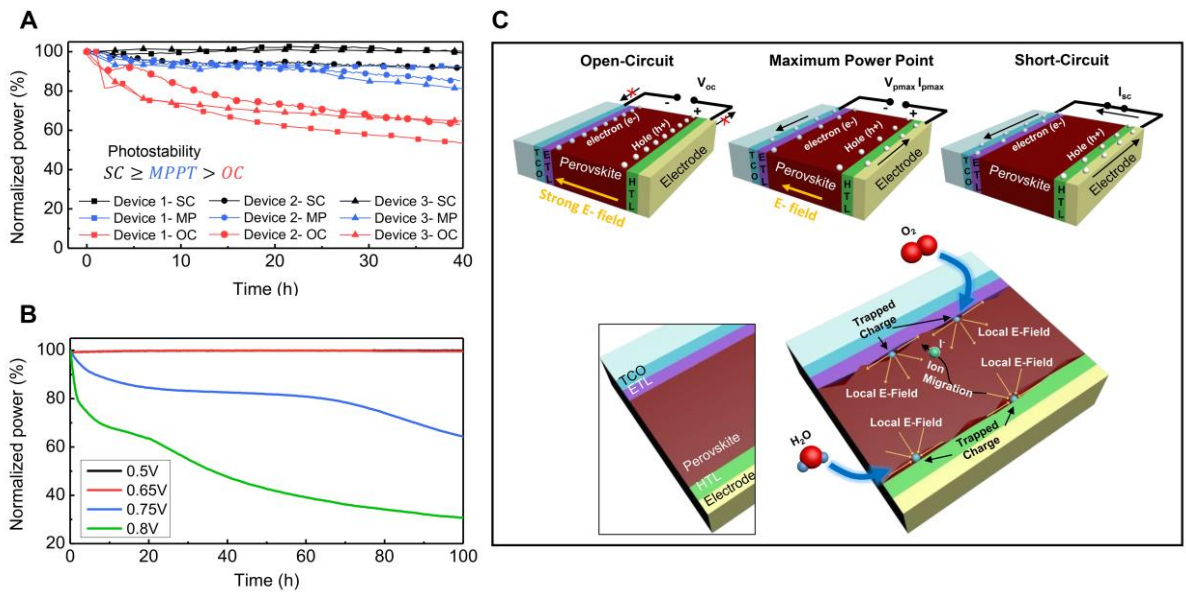

**Fig. 1. Bias-dependent degradation rates of perovskite solar cells** A) Time evolution of the normalized power of three types of devices measured at SC(short circuit), MPPT(maximum power point tracking) and OC(open circuit) condition under one sun illumination in ambient condition(RH=30%). All devices were glass-encapsulated. B) Time evolution of the normalized power of device 3 measured at four different voltage biases. 0.5V(black), 0.65V(red), 0.75V(blue), 0.8V(green). C) A schematic illustration describing the effect of load bias on charge accumulation, ion migration and resulting degradation.

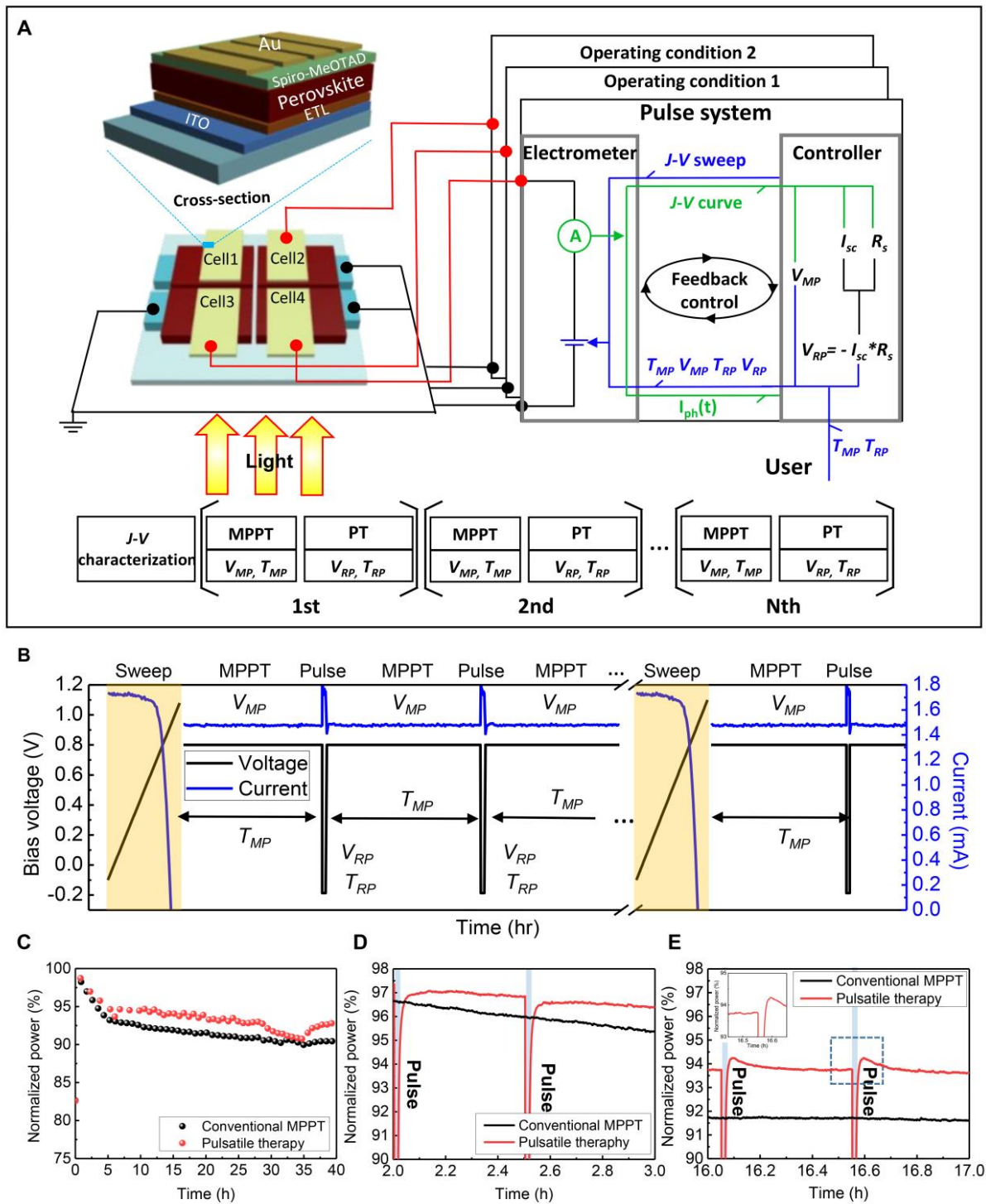

**Fig. 2. Pulsatile therapy for perovskite solar cells** A) A schematic illustration of our novel pulsed-MPPT system. B) The profile of applied bias voltage and corresponding photocurrent over time in our pulsed-MPPT system. $V_{MP}$: Maximum power point voltage, $T_{MP}$: Time duration for maximum power point voltage, $V_{RP}$: Reverse pulse voltage, $T_{RP}$: Time duration for reverse pulse voltage. C)Time evolution of normalized power for PT(pulsatile therapy)- and MPPT(Maximum power point tracking)- tested cells. Comparison of normalized power for PT- and MPPT- tested cells after D)2h of operation and E)16h of operation. All devices were glass-encapsulated.

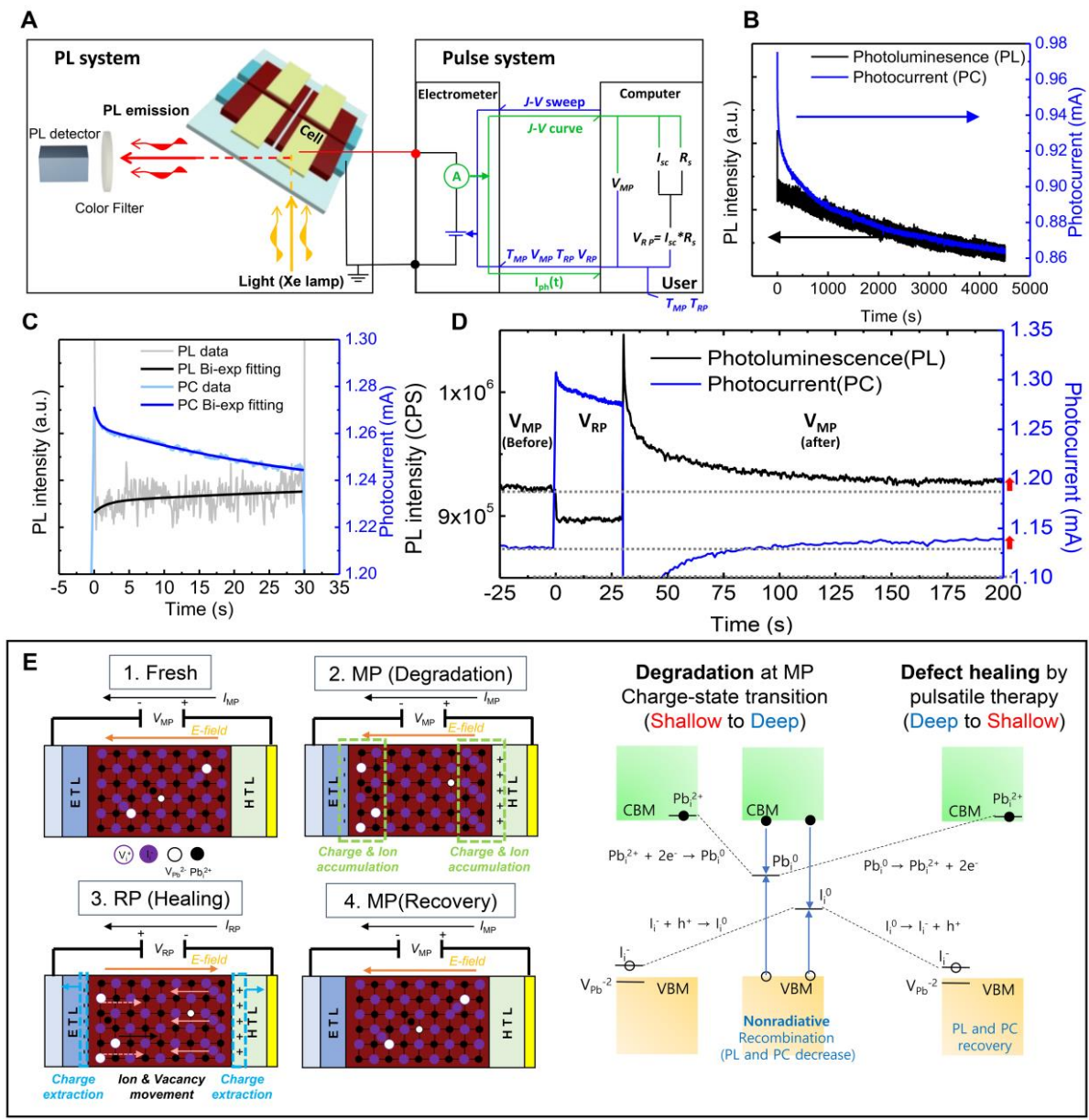

**Fig. 3. Defect healing mechanism underlying pulsatile therapy** A) A schematic illustration of kinetic photoluminescence measurement with PT system. B) PC(photocurrent) and PL(photoluminescence) kinetics of the device operated at MP condition under continuous light illumination. C) Kinetics of PL intensities and PC during RP (30 sec) applied after MP (30 min) operation. D) Overall kinetics of PL intensity and PC during one cycle of pulsatile therapy($V_{MP}$-$V_{RP}$-$V_{MP}$). E) A Schematic illustration for the mechanism underlying PT. (left) spatial charge and ionic defect distribution during PT (right) defect dynamics based on charge-state transition during PT.

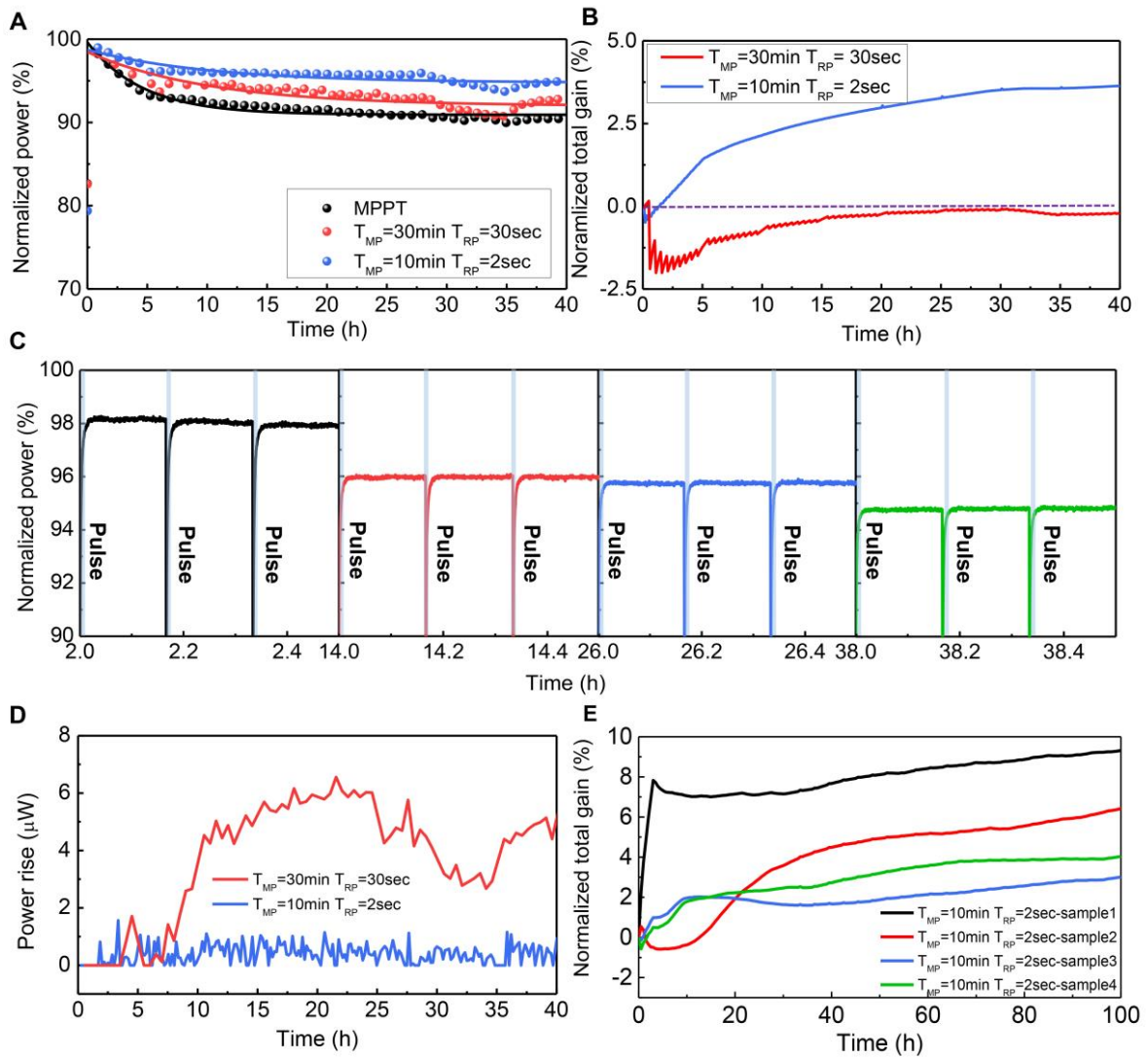

**Fig. 4. Optimized pulsatile therapy for maintaining defect-tolerance** A) Time evolution of the normalized power of the devices operated by MPPT (black dot), $T_{MP}$=30min $T_{RP}$=30sec (red dot), and $T_{MP}$=10min $T_{RP}$=2sec (blue dot) condition under one sun illumination in ambient condition (RH=30%). B) Normalized total gain of devices tested by $T_{MP}$=30min $T_{RP}$=30sec and $T_{MP}$=10min $T_{RP}$=2sec condition compared to MPPT. C) Normalized power profile of the device tested by $T_{MP}$=10min $T_{RP}$=2sec condition after 2h(black), 14hr(red), 26hr(blue), 38hr(green). D) Time evolution of power rise of the devices operated by $T_{MP}$=30min $T_{RP}$=30sec, and $T_{MP}$=10min $T_{RP}$=2sec condition E) Normalized total gain of four samples operated by $T_{MP}$=10min $T_{RP}$=2sec condition for 100hr light illumination.

|  | Reverse pulse | | MP after reverse pulse | |
|---|---|---|---|---|
| Time constant | T1 (s) | T2 (s) | T1 (s) | T2 (s) |
| Photocurrent (PC) | 0.593 | 34.23 | 0.434 | 7.12 |
| Photoluminescence (PL) | 0.635 | 40.7 | 0.745 | 22.7 |

**Table 1. Photocurrent dynamics during pulsatile therapy** Time constant extracted from kinetics of PL intensities and PC during RP (30 sec) and MP (30min) operation after RP, respectively. In RP, PC decay and PL increase was fitted by bi-exponential fitting. the fast one resulted from capacitive current and the slow one resulted from field-induced ion movement

# Supplementary Materials for

# Pulsatile therapy for perovskite solar cells

*Kiwan Jeong[1†], Junseop Byeon[1,2†], Jihun Jang[1], Namyoung Ahn[1]\*,and Mansoo Choi[1,2]\**

† These authors contributed equally to this work

\*Corresponding author E-mail: mchoi@snu.ac.kr, nyny92@snu.ac.kr

**This material includes:**

    Materials and Methods

    Supplementary Text

    Figs. S1 to S12

.

Materials and Methods

*Fabrication of perovskite solar cells*

The Indium-doped tin oxide (ITO) glass substrate were pre-patterned by 532nm pico-second laser to fabricate four electrically-independent cells in the device. The laser power was 1.9W and the scribed line width was 300um. Patterned ITO glass substrates (AMG, 9.5Ωcm$^{-2}$) were cleaned by sonication sequentially using acetone, isopropanol, and deionized water. For the MAPbI$_3$- based perovskite solar cells (PSCs) with C$_{60}$ as the electron transporting layer (device 1), a C$_{60}$ layer (35 nm) was deposited on the cleaned ITO glass substrate using the vacuum thermal evaporator at deposition rate of 0.2 Å s$^{-1}$. For MAPbI$_3$-PSCs with SnO$_2$ as the ETL(device 2), SnO$_2$ layer was fabricated on the ITO glass substrate by spin-coating 2.67wt% of SnO$_2$ colloid precursor (tin(IV) oxide, 15% in H2O colloidal dispersion, Alfa Aesar) in DI water at 4000 rpm for 30 s. The SnO$_2$ layer was annealed at 150 °C for 30 min. The thickness of the SnO$_2$ layer was around 30 nm. A precursor solutions of MAPbI$_3$ were prepared by adding 461 mg of PbI$_2$ (Alfa Aesar) and 159 mg of MAI (Great solar) and 78 mg of mixed adducts dimethyl sulfoxide (DMSO; Sigma-Aldrich) with 5mol% urea in 0.55 mL of *N,N*-dimethylformamide (DMF; Sigma-Aldrich). The solution was spin-coated on the ETL layer at 4,000 rpm for 20 s with 0.5 mL of diethyl ether dripping treatment. The film was annealed at 115 °C for 20 min. For the Triple perovskite- based PSCs with SnO$_2$ as ETLs (device 3), the triple perovskite was deposited by 2-step spin coating method. First, 1.25 M of PbI$_2$ with 5mol% of CsCl in 0.05ml of DMSO and 0.95ml of DMF was spin coated onto the ETL at 2,500 rpm for 30 s. The mixture solution of FAI:MABr:MACl (75 mg:7.5 mg: 7.5 mg in 1 ml isopropanol) was spin coated onto the CsCl/PbI$_2$ film at 5,000 rpm for 30 s, then annealed at 150 ◦C for 20 min. In order to prepare a solution for the hole-transporting layer (HTL), 72.3 mg of spiro-MeOTAD (Merck) was dissolved in 1 mL of chlorobenzene (Sigma-Aldrich). 28.8 μL of 4-tert-butyl pyridine and 17.5 μL of lithium bis(trifluoromethanesulfonyl)imide from a stock solution (520 mg of lithium bis(trifluoromethanesulfonyl)imide in 1 mL of acetonitrile, 99.8% purity, Sigma-Aldrich) were added to the mixture solution. The HTL was formed on the perovskite film by spin-coating mixture solution at 2,500 rpm for 30 s. A gold layer with a thickness of 50 nm was deposited on the HTL by using the vacuum thermal evaporator at deposition rate of 0.3 Å s$^{-1}$. All spin-coating processes were carried out in a dry room (<15% relative humidity, at room temperature). Fabricated solar cells were encapsulated with glass using UV cured resin (XNR5570, NAGASE) in glove box.

*Characterization*

**J-V measurement**

The current–voltage characteristics were measured by a solar simulator (Sol3A, Oriel) and a source-meter (2400, Keithley) under AM 1.5G at 100 mW cm$^{−2}$ at room temperature inside a glove box. The light intensity was calibrated by using a Si reference cell (Rc-1000-TC-KG5-N, VLSI Standards, USA). The aperture size of PSCs is 0.0729 cm$^2$.

**Long-term stability test with PT**

The current–voltage characteristics for aging under 1-sun light illumination were measured by a solar simulator (K3000, McScience) and a source-meter (2400, Keithley). Customized JIG was designed to contact cathode/anode of each cell on each independent section of etched ITO, independently. LABView software was used to design pulsed MPPT tracking system by controlling source-meter via GPIB. The system was set to perform *J-V* sweep periodically (3hr or 5hr were selected), with scan rate of 0.06 V/s, scan range of -0.1 V~1.1 V, voltage step of 0.04 V, and both reverse/forward direction. Recent parameters for feedback ($I_{sc}$, $V_{oc}$, $R_{sh}$, $R_s$, $FF$) were automatically updated and calculated based on averaged values of *J-V* curve for both reverse and forward direction. $T_{MP}$ and $T_P$ were set prior to system operation, while $V_{RP}$ ($-I_{sc}*R_s$) and $V_{MP}$ were calculated by updated parameters from recently measured *J-V* sweep. The system also measured and stored realtime photocurrent values, of which the sampling rate was 0.2Hz during MPPT, and 10Hz during pulse. If a *J-V* sweep and a pulse overlap, *J-V* sweep was set as the highest hierarchy.

### Kinetic photoluminescence measurement

Steady-state and kinetic photoluminescence (PL) measurements were conducted using a FluoroMax-4 spectrofluorometer (Horiba). Xenon lamp(150W) was used as a light source with wavelength near 463 nm selected using spectroscope. Voltage can be applied simultaneously by wiring device to K2400. The resulting PL was measured using high-sensitivity photodetector targeting wavelength of 780 nm.

### Charge extraction measurement

In order to measure the accumulated charges in the PSCs, charge extraction was conducted, and Figure S# shows the detailed procedure. We carried out the charge extraction with delay times ranging from 0.5 ms, and a cluster of white LEDs with a power density of 100 mW cm$^{-2}$ was used as the light source. The charge extraction measurement was carried out using an electrochemical workstation (Autolab 320N, Metrohm, Switzerland) with an Autolab LED Driver Kit (Metrohm, Switzerland). White LEDs with a power density of 100 mW cm$^{-2}$ was used as the light source.

**Supplementary Note 1**

Equivalent circuit model

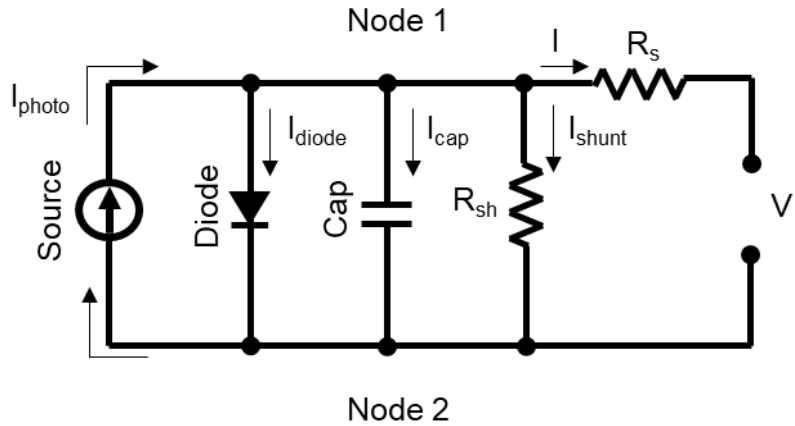

We simply modeled the equivalent circuit of perovskite solar cells by adding capacitance parallel to shunt resistance. This circuit is for the DC voltage bias, which reflects capacitive characteristics of perovskite solar cells. To find an appropriate reverse bias for effective charge extraction, we calculated the external bias V for no electrical potential difference between node 1 and 2 by solving Kirchhoff's law of the equivalent circuit. The electrical potential between node 1 and 2 can be expressed as below.

$$\Delta V_{12} = IR_S + V$$

For no electrical potential difference,

$$IR_S + V = 0, V = -IR_S$$

Photocurrent at a reverse bias is similar to short-circuit current. We can assume

$$I \approx I_{sc}$$

Finally, we can obtain the value of the reverse bias voltage for no potential difference

$$V = -I_{sc}R_S$$

The reverse bias can extract charges by minimizing the effect of capacitance in the circuit model. Also, the value is not large, which makes the energy loss very low.

**Supplementary Note 2**

Normalized total gain

To estimate the improvement of total yield by pulsatile therapy compared to MPPT, we obtained normalized power as a function of time. The normalized power $P_{norm}$ is calculated by

$$P_{norm}(t) = \frac{P(t)}{P(0)}$$

where P(t) is an electrical power harvested by a device at operation time t and P(0) is an initial electrical power. P(t) has negative values for reverse biases and positive ones for maximum power point forward biases in the present pulsatile therapy.
We calculated integration of normalized power over time to estimate total yield during total operation time as the following equation.

$$Y_{norm}(T) = \int_0^T P_{norm}(t)dt$$

To calculate how much the total yield of therapy has improved over the total yield of the MPPT case, we defined normalized total gain (G) as below.

$$G(\%) = 100 * (Y_{norm,PT} - Y_{norm,MPPT})/Y_{norm,MPPT}$$

In this calculation method, when G(%) is equal to zero, total yield of the therapy is the same to that of the MPPT. The normalized total gain was always calculated based on the MPPT case which was simultaneously performed with the therapy.

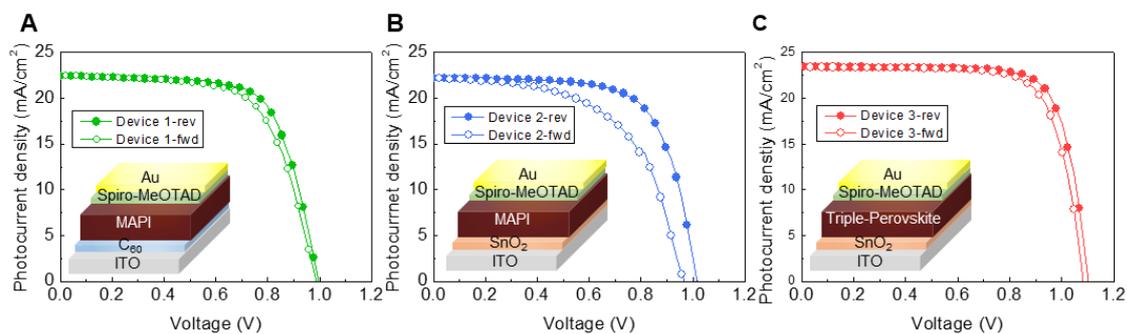

| | $V_{oc}$ (V) | $J_{sc}$ (mA/cm$^2$) | FF(%) | PCE(%) |
|---|---|---|---|---|
| Device 1-rev | 1.00 | 22.4 | 69.0 | 15.4 |
| Device 1-fwd | 0.99 | 22.4 | 65.0 | 14.4 |
| Device 2-rev | 1.01 | 22.1 | 68.6 | 15.4 |
| Device 2-fwd | 0.96 | 22.1 | 56.7 | 12.1 |
| Device 3-rev | 1.10 | 23.4 | 76.6 | 19.8 |
| Device 3-fwd | 1.08 | 23.4 | 74.2 | 18.8 |

Fig. S1. *J-V* characteristic of the A) device 1 (ITO/C$_{60}$/MAPbI$_3$/HTL/Au), B) device 2 (ITO/SnO$_2$/MAPbI$_3$/HTL/Au), C) device 3 (ITO/SnO$_2$/Triple-perovskite/HTL/Au) measured in reverse (full circle) and forward (hollow circle) scan.

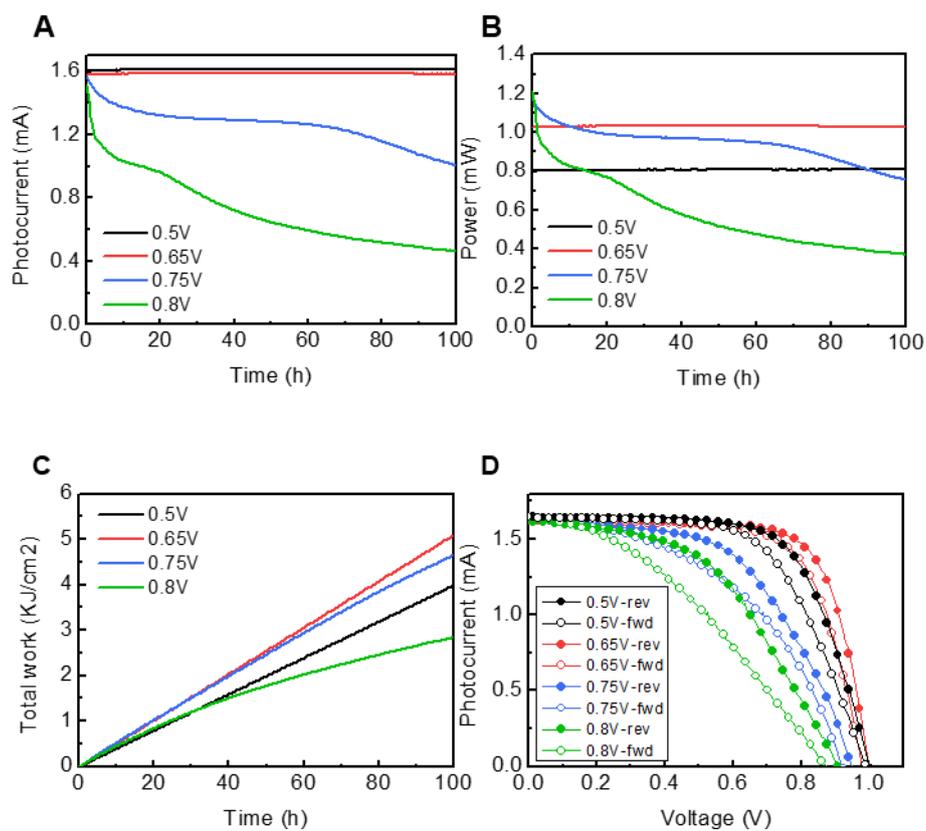

Fig. S2. Time evolution of the normalized power of device 3 measured at four different voltage biases. 0.5V(black),0.65V(red), 0.75V(blue), 0.8V(green). A) photocurrent, B) power and C) Total work, D) *I-V* characteristic of 100hr aged device measured in reverse(full circle) and forward(hollow circle) scan. Note that device stability was better as load bias was smaller.

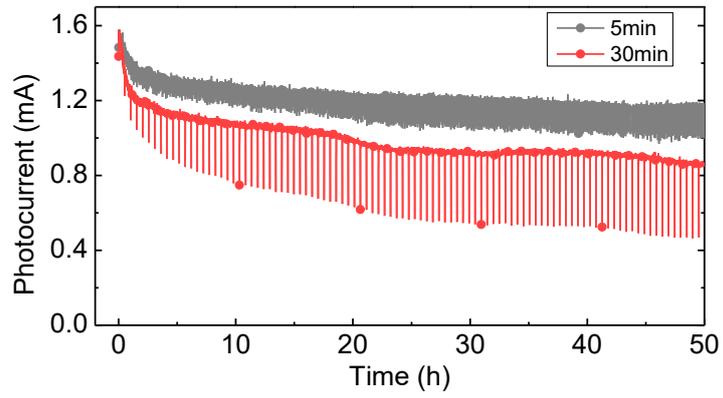

Fig. S3. Time evolution of photocurrent with different interval of *I-V* characterization under one sun at 0.7V. Note that shorter *I-V* characterization interval shows better stability.

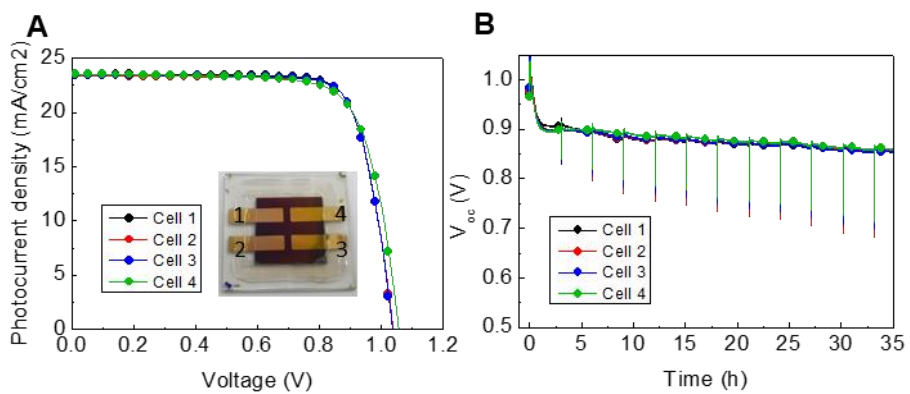

| Device3 | $V_{oc}$ (V) | $J_{sc}$ (mA/cm$^2$) | FF(%) | PCE(%) |
|---|---|---|---|---|
| Cell1-rev | 1.06 | 22.9 | 73.9 | 18.0 |
| Cell2-rev | 1.05 | 23.5 | 72.6 | 17.9 |
| Cell3-rev | 1.07 | 22.6 | 70.8 | 17.2 |
| Cell4-rev | 1.04 | 22.9 | 75.5 | 17.9 |

Fig. S4. A) *J-V* characteristic of 4 cells in device 3, reverse scan. b) Time evolution of $V_{oc}$ of each cell in device3.

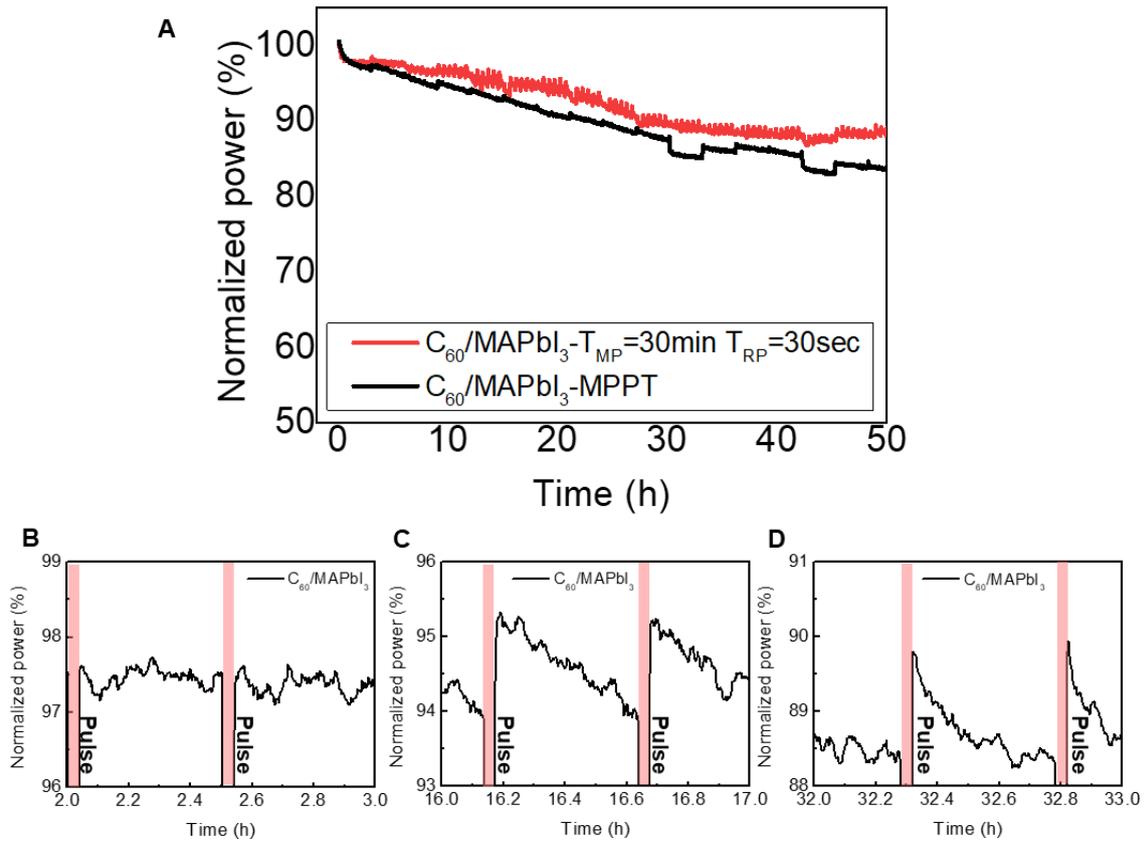

Fig. S5. Time-evolving normalized power of A) $C_{60}$/MAPbI$_3$-device operated by PT (pulsatile therapy)- and MPPT (Maximum power point tracking) condition. Normalized power profile of $C_{60}$/MAPbI$_3$ device operated by $T_{MP}$=30min $T_{RP}$=30sec condition after B) 2h, C) 16h and D) 32h under one sun illumination. Magnitude of PC rise was getting larger in later time.

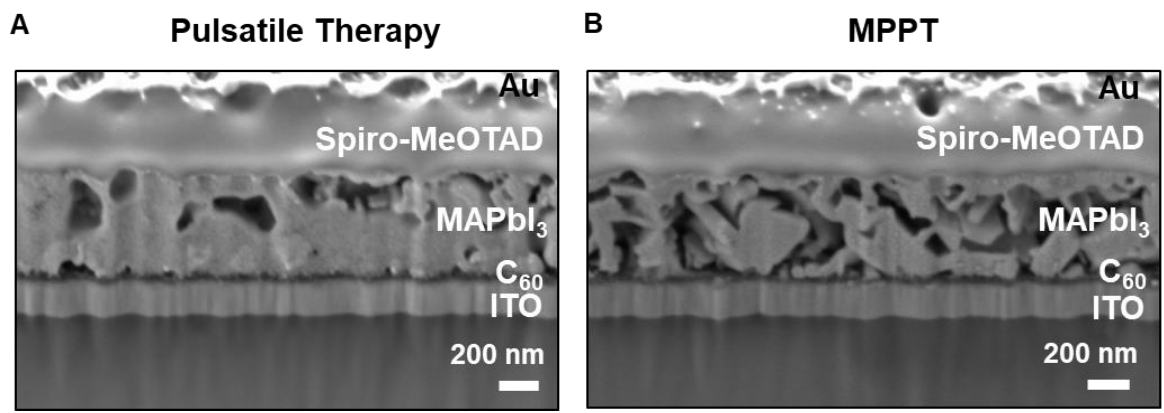

Fig. S6. Cross-sectional SEM images of devices operated by A) pulsatile therapy (PT) and B) MPPT after the light induced degradation

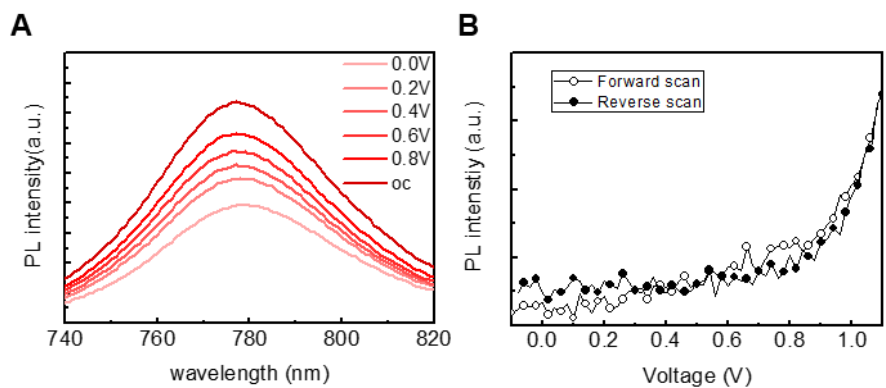

Fig. S7. A) Steady-state PL under different bias voltage. B) *PL-V* curve of device measured in reverse (full circle) and forward (hollow circle) scan at PL peak wavelength.

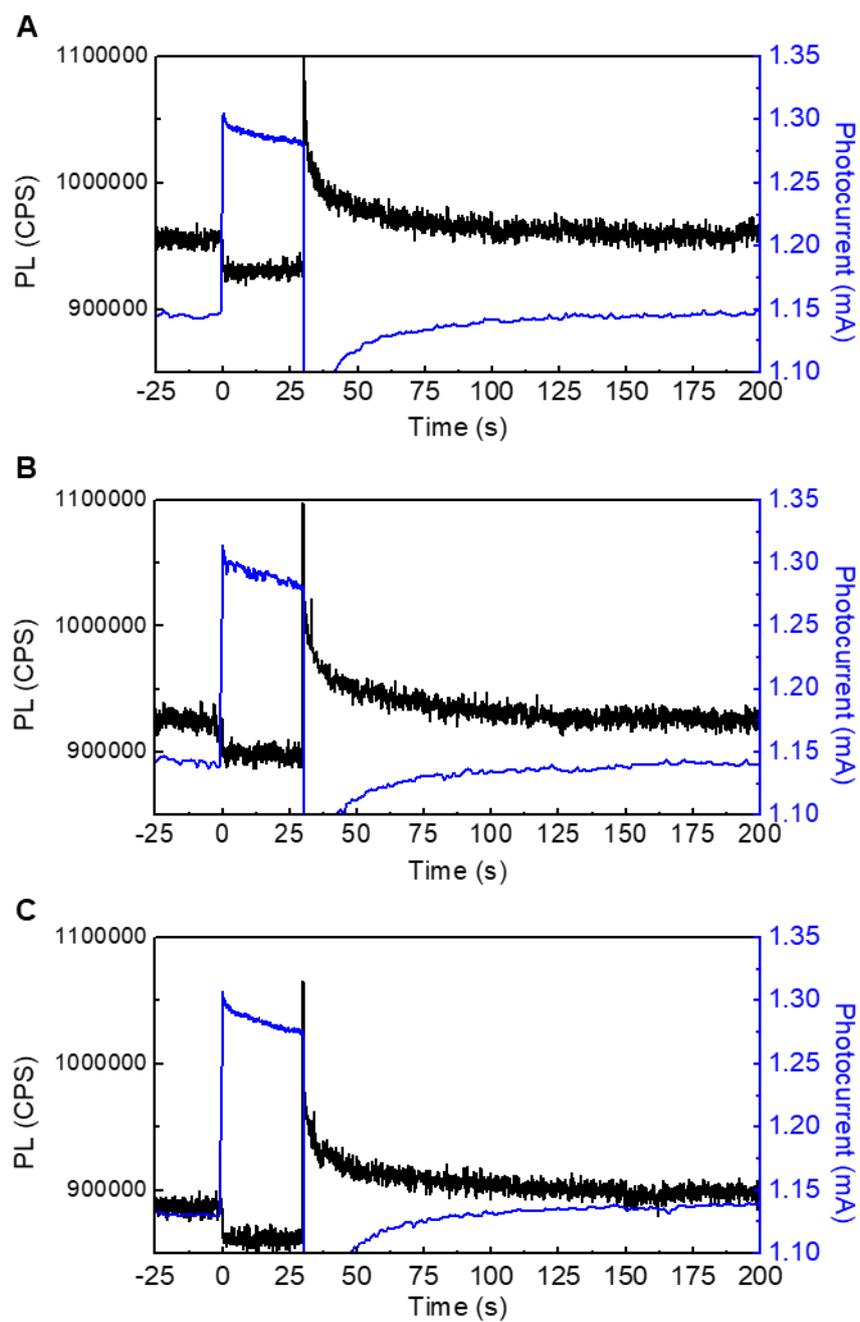

Fig. S8. Kinetics of transient photocurrent and PL(at PL peak wavelength) during $T_{MP}$=30min $T_{RP}$=30sec PT cycle. a)1st cycle, b)2nd cycle and c)3rd cycle.

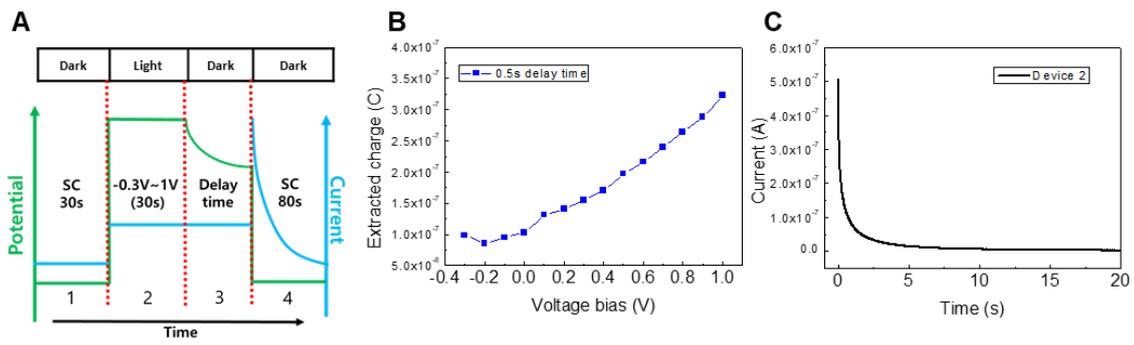

Fig. S9. A) Process of charge extraction measurement. B) Quantity of extracted charges as a function bias voltage for $SnO_2$/$MAPbI_3$-devices (Device 2). Delay time was 0.5s C) Current profile of $SnO_2$/$MAPbI_3$-device at 4$^{th}$ step in charge extraction measurement. Current was extracted within 5s.

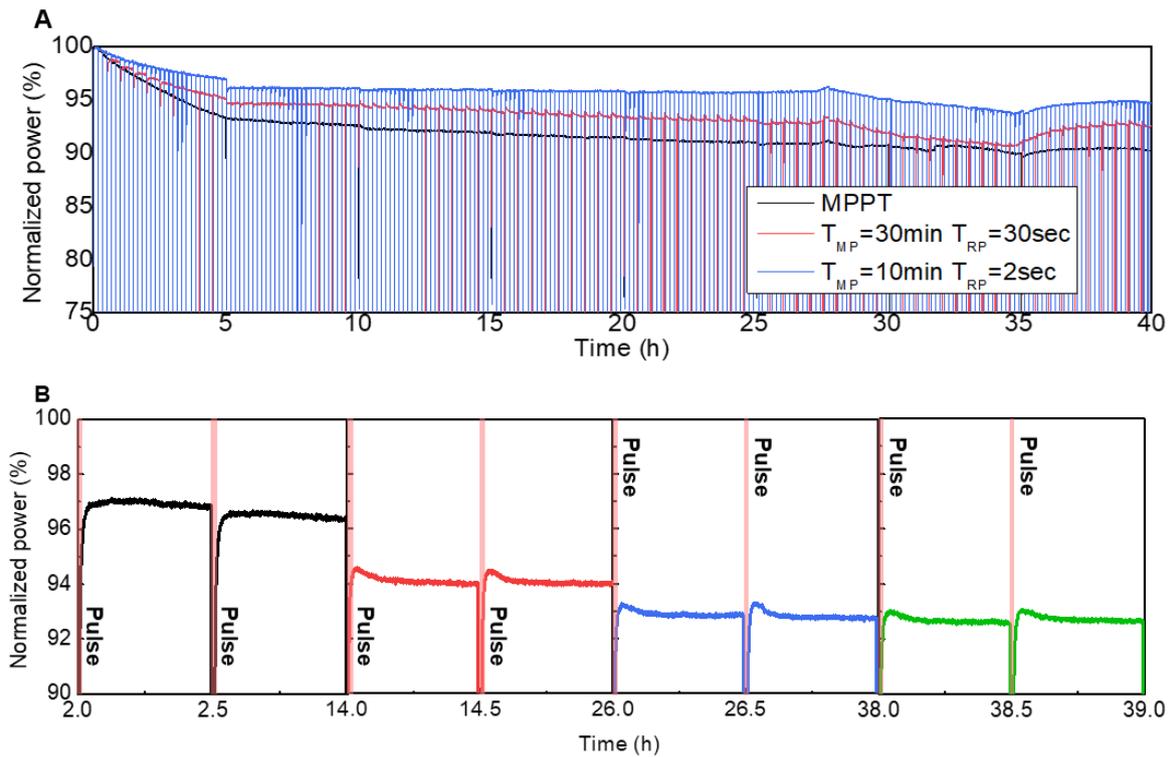

Fig. S10. A) Time dependent normalized power of devices operated by $T_{MP}$=10min $T_{RP}$=2sec, and $T_{MP}$=30min $T_{RP}$=30sec condition including all transient signals during pulsatile therapy. Transient increase of photocurrent right after RP→MP or *J-V* characterization all included in raw data. B) Normalized power profile of device 3 in $T_{MP}$=30min $T_{RP}$=30sec condition after 2h(black), 14hr(red), 26hr(blue), 38hr(green).

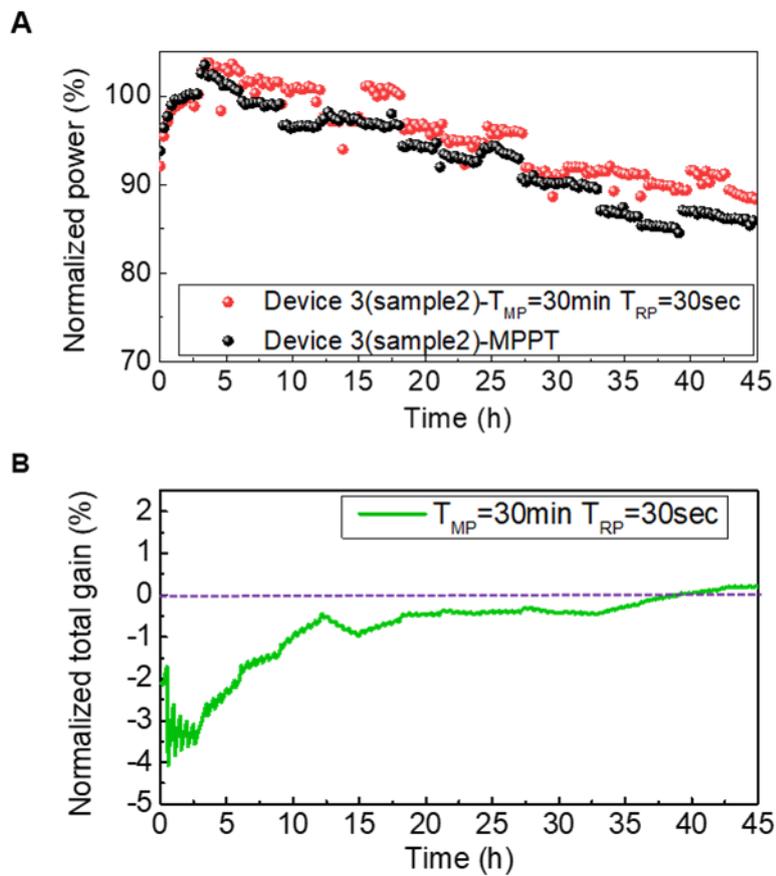

Fig. S11. A) Time dependent normalized power of device 3 operated by MPPT, and TMP=30min TRP=30sec condition including transient signal due to pulsatile therapy. B) Normalized total gain of device 3 operated by TMP=30min TRP=30sec condition. The gain exceeded the threshold (0%) after 40hr operation under TMP=30min TRP=30sec condition. The devices were operated under one sun illumination.

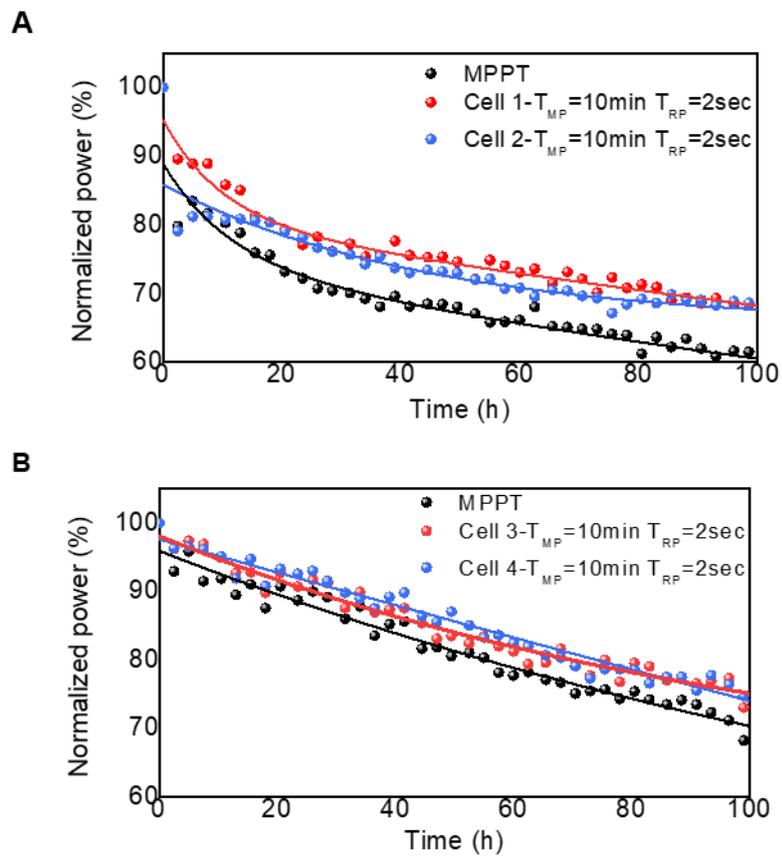

Fig. S12. Time evolution of normalized power of four samples (SnO$_2$/FA-rich perovskite/Spiro-MeOTAD) with same device structure operated T$_{MP}$=10min T$_{RP}$=2sec condition until 100hr